\documentclass[12pt]{iopart}

\usepackage{epsfig}
\usepackage{graphicx}
\usepackage{bm}

\begin{document}
\title{An effective chiral Hadron-Quark Equation of State}
%
%
\author{J.~Steinheimer}
\address{Institut f\"ur Theoretische Physik, Goethe-Universit\"at, Max-von-Laue-Str.~1,
D-60438 Frankfurt am Main, Germany}
\author{S.~Schramm}
\address{Institut f\"ur Theoretische Physik, Goethe-Universit\"at, Max-von-Laue-Str.~1,
D-60438 Frankfurt am Main, Germany}
\address{Center for Scientific Computing, Max-von-Laue-Str.~1, D-60438 Frankfurt am Main}
\address{Frankfurt Institute for Advanced Studies (FIAS), Ruth-Moufang-Str.~1, D-60438 Frankfurt am Main,
Germany}

\author{H.~St\"ocker}
\address{Institut f\"ur Theoretische Physik, Goethe-Universit\"at, Max-von-Laue-Str.~1,
D-60438 Frankfurt am Main, Germany}
\address{Frankfurt Institute for Advanced Studies (FIAS), Ruth-Moufang-Str.~1, D-60438 Frankfurt am Main,
Germany}
\address{GSI Helmholtzzentrum f\"ur Schwerionenforschung GmbH, Planckstr.~1, D-64291 Darmstadt, Germany}

\begin{abstract}
We construct an effective model for the QCD equation of state, taking
into account chiral symmetry restoration as well as the deconfinement phase transition.
The correct asymptotic degrees of freedom at the high and low temperature limits are
included (quarks $\leftrightarrow$ hadrons). The model shows a rapid crossover for both order parameters, as is expected from lattice calculations. We then compare the thermodynamic properties of
the model at $\mu_B=0$ which turn out to be in qualitative agreement with lattice data, while apparent quantitative differences can be attributed to hadronic contributions and excluded volume corrections.
Furthermore we discuss the effects of a repulsive vector type quark interaction at finite baryon number densities on the resulting phase diagram of the model. Our current model is able to reproduce a first-order liquid gas phase transition as expected, but does not show any signs of a first order deconfinement or chiral phase transition. Both transitions rather appear as a very wide crossover in which heavily medium modified hadron coexist with free quarks. 
\end{abstract}


\maketitle

\section{Introduction}
The recent experimental results at the Relativistic Heavy Ion Collider (RHIC), suggesting the creation of a nearly perfect fluid
\cite{Adams:2005dq,Back:2004je,Arsene:2004fa,Adcox:2004mh},
 have fueled interest in the study of bulk properties of strongly interacting matter (QCD).
 Heavy ion experiments at different beam energies try to map out the QCD phase diagram, especially
 the region where one expects a phase transition from a confined gas of hadrons to a deconfined state of
 quarks and gluons (QGP)
 \cite{Gyulassy:2004zy,Fodor:2001pe,Fodor:2007vv,Karsch:2004wd,Stephanov:1998dy,Gazdzicki:1998vd,Stephanov:1999zu,Bravina:1999dh,Bravina:2000dk,Gazdzicki:2004ef,Arsene:2006vf,Theis:1984qc}. To relate any experimental observables to the properties of the matter produced in heavy ion collisions, a deeper understanding of the thermodynamics of QCD has to be obtained and integrated in model simulations of these collisions.\\
One can define two different phase transitions, the first being the chiral phase transition associated with chiral symmetry restoration in the vanishing quark mass limit, where the chiral condensate serves as a well defined order parameter. In the limit of heavy quarks, a deconfinement phase transition with the Polyakov loop as order parameter, is assumed. Physical quarks however have intermediate masses and one would expect that at least the deconfinement order parameter is not so well defined anymore. Both transitions seem to be correlated, accounting for (some) lattice QCD observation that deconfinement and chiral restoration occur at the same temperature (at least at $\mu_B=0$) \cite{Fukugita:1986rr,Karsch:1994hm,Aoki:1998wg,Karsch:2000kv,Allton:2002zi}, while newest results indicate at least a small shift in the critical temperatures \cite{Borsanyi:2010cj,Borsanyi:2010bp}.\\  
The lattice calculations at finite temperature are an important tool for the investigation of the QCD phase diagram.
For the thermodynamics of the pure gauge theory high accuracy data is available \cite{Boyd:1996bx}, and the equation of state (EoS) of strongly interacting matter at vanishing chemical potential is reasonably well understood \cite{Karsch:2000ps,Aoki:2006we}.
Here lattice predicts a rapid crossover for the deconfining and chiral phase transitions.\\
At finite baryo-chemical potential, lattice calculations suffer from the so-called sign problem. There are several
different approaches to obtain results at
finite $\mu_B$ \cite{Fodor:2002km,Allton:2002zi,Allton:2003vx,Allton:2005gk,deForcrand:2003hx,Laermann:2003cv,DElia:2004at,DElia:2002gd}, but yet no clear picture, especially about the existence and location of a possible critical end point, has emerged.\\
Recent considerations based on connecting the large $N_c$ limit with real-word QCD
draw an even more exotic picture of the phase diagram,
where the critical temperatures of the deconfinement and chiral phase transitions disconnect and depart
in the region of high net baryon densities \cite{Hidaka:2008yy}.\\

PNJL-type models have recently been used to successfully describe lattice results on bulk properties of a strongly interacting matter \cite{Fukushima:2003fw,Ratti:2005jh} (see also \cite{Ratti:2004ra,Roessner:2006xn,Sasaki:2006ww,Ratti:2007jf,Rossner:2007ik,Ciminale:2007sr,Schaefer:2007pw,Fu:2007xc,Hell:2008cc,Abuki:2008nm,Fukushima:2008wg,Fukushima:2008is,Costa:2008gr,Costa:2008dp,Hansen:2006ee,Mukherjee:2006hq,Abuki:2008tx,Abuki:2008ht,Fukushima:2009dx,Mao:2009aq,Schaefer:2009ui}). These constituent quark models have the correct degrees of freedom in the asymptotic regime of free quarks and gluons but lack the rich hadronic spectrum.

In our approach we combine, in a single model, a well-established flavor-SU(3) hadronic model with a PNJL-type quark-gluon description of the highly excited matter. This allows us to study the chiral-symmetry and confinement-deconfinement phase structure of the strongly interacting matter at high temperatures and densities. In addition we obtain an equation of state of hadronic and quark matter
that is applicable over a wide range of thermodynamical conditions and that can therefore be used in heavy-ion simulations with
very different beam energies.

\section{Model description}
In our approach we derive the EoS of hot and dense nuclear matter using a single model for the hadronic and quark phase (later also referred to as the hadron quark model HQM).
The model includes the correct asymptotic degrees of freedom, namely a free gas of quarks and gluons at
infinite temperature, and a gas of hadrons having the correct vacuum properties at vanishing temperature. The model also predicts the structure of finite nuclei, nuclear and neutron matter properties and a first order liquid-vapor phase transition.
The two phase transitions that are expected from QCD, the chiral and deconfinement transitions, are included in a consistent manner.\\
In the following we will show how we describe the different phases of QCD and how we combine them in a unified approach.\\

We describe the hadronic part of the EoS, using a flavor-SU(3) model which
is an extension of a non-linear representation of a sigma-omega model including
the pseudo-scalar and vector octets of mesons and the baryonic octet and decuplet
(for a detailed discussion see \cite{Papazoglou:1998vr,Papazoglou:1997uw,Dexheimer:2008ax}).

The Lagrangian density of the model in mean field approximation reads:
\begin{eqnarray}
&L = L_{kin}+L_{int}+L_{meson},&
\end{eqnarray}
where besides the kinetic energy term for hadrons, the terms:

\begin{eqnarray}
&L_{int}=-\sum_i \bar{\psi_i}[\gamma_0(g_{i\omega}\omega+g_{i\phi}\phi)+m_i^*]\psi_i,&
\label{formel1}
\end{eqnarray}

\begin{eqnarray}
&L_{meson}=-\frac{1}{2}(m_\omega^2 \omega^2+m_\phi^2\phi^2)\nonumber&\\
&-g_4\left(\omega^4+\frac{\phi^4}{4}+3\omega^2\phi^2+\frac{4\omega^3\phi}{\sqrt{2}}+\frac{2\omega\phi^3}{\sqrt{2}}\right)\nonumber&\\
&+\frac{1}{2}k_0(\sigma^2+\zeta^2)-k_1(\sigma^2+\zeta^2)^2&\nonumber\\
&-k_2\left(\frac{\sigma^4}{2}+\zeta^4\right)-k_3\sigma^2\zeta&\nonumber\\
&+ m_\pi^2 f_\pi\sigma&\nonumber\\
&+\left(\sqrt{2}m_k^ 2f_k-\frac{1}{\sqrt{2}}m_\pi^ 2 f_\pi\right)\zeta~,&\nonumber\\
&+ \chi^4-\chi_0^4 + \ln\frac{\chi^4}{\chi_0^4} -k_4\ \frac{\chi^4}{\chi_0^4} \ln{\frac{\sigma^2\zeta}{\sigma_0^2\zeta_0}} ~.&
\label{formel2}
\end{eqnarray}
represent the interactions between baryons
and vector and scalar mesons, the self-interactions of
scalar and vector mesons, and an explicitly chiral symmetry breaking term.
The index $i$ denotes the baryon octet and decuplet. Here, the mesonic condensates (determined in
mean-field approximation) included are
the vector-isoscalars $\omega$ and $\phi$, and
the scalar-isoscalars $\sigma$ and $\zeta$ (strange quark-antiquark state). Assuming isospin symmetric matter, we neglected the $\rho$-meson contribution in Eq. \ref{formel2}.

The concept of a broken scale invariance leads to the trace anomaly in (massless) QCD.
The last four terms of (\ref{formel2}) were introduced to mimick this anomaly on tree level \cite{Schechter:1980ak,Papazoglou:1997uw}. The effect of the logarithmic term $ \chi^4 \ln{\chi}$ is to break the scale invariance which leads to the proportionality
$\chi^4 \propto \theta^{\mu}_{\mu}$.
The comparison of the trace anomaly of QCD with that of the effective theory allows for the identification of the
$\chi$ field with the gluon condensate $\theta^{\mu}_{\mu}$. This holds only, if the meson-meson potential is scale invariant and can be achieved by multiplying the invariants of scale dimension less then four with an appropriate power of the dilaton field $\chi$.

The effective masses of the baryons (of the octet)
are generated by the scalar mesons except for an explicit
mass term ($\delta m_N=120$ MeV):
\begin{eqnarray}
&m_{b}^*=g_{b\sigma}\sigma+g_{b\zeta}\zeta+\delta m_b,&
\end{eqnarray}
while, for simplicity and in order to reduce the number of free parameters, the masses of the decuplet baryons are kept at their vacuum expectation values.
With the increase of temperature/density, the $\sigma$ field (non-strange chiral condensate) decreases
in value, causing the effective masses of the particles to decrease towards chiral symmetry restoration.
The coupling constants for the baryons \cite{Dexheimer:2009hi} are chosen
to reproduce the vacuum masses of the baryons, nuclear saturation properties and
asymmetry energy as well as the $\Lambda$-hyperon potentials. The vacuum expectation values of the scalar
mesons are constrained by reproducing the pion and kaon decay constants.\\

The extension of the hadronic SU(3) model to quark degrees of freedom
is constructed in analogy to the PNJL model.
The sigma model uses the Polyakov loop $\Phi$ as the order parameter for
deconfinement. $\Phi$ is defined via $\Phi=\frac13$Tr$[\exp{(i\int d\tau A_4)}]$, where $A_4=iA_0$ is the temporal component
of the SU(3) gauge field. One should note that one must distinguish $\Phi$, and its conjugate $\Phi^{*}$
at finite baryon densities \cite{Fukushima:2006uv,Allton:2002zi,Dumitru:2005ng}, as they couple differently to quarks and antiquarks respectively.\\
In our approach the effective masses of the quarks
are generated by the scalar mesons except for a small explicit
mass term ($\delta m_q=5$ MeV and $\delta m_s=105$ MeV for the strange quark):
\begin{eqnarray}
&m_{q}^*=g_{q\sigma}\sigma+\delta m_q,&\nonumber\\
&m_{s}^*=g_{s\zeta}\zeta+\delta m_s,&
\end{eqnarray}
with values of $g_{q\sigma}=g_{s\zeta}= 4.0$.
Vector type interactions introduce an effective chemical potential for the quarks and baryons, generated by the coupling to the vector mesons:
$\mu^*_i=\mu_i-g_{i\omega}\omega-g_{i\phi}\phi$.\\
A coupling of the quarks to the Polyakov loop is introduced in the thermal energy of the quarks. Their thermal contribution to the grand canonical potential $\Omega$, can then be written as:
\begin{equation}
	\Omega_{q}=-T \sum_{i\in Q}{\frac{\gamma_i}{(2 \pi)^3}\int{d^3k \ln\left(1+\Phi \exp{\frac{E_i^*-\mu^{*}_i}{T}}\right)}}
\end{equation}
and
\begin{equation}
	\Omega_{\overline{q}}=-T \sum_{i\in Q}{\frac{\gamma_i}{(2 \pi)^3}\int{d^3k \ln\left(1+\Phi^* \exp{\frac{E_i^*+\mu^{*}_i}{T}}\right)}}
\end{equation}

The sums run over all quark flavors, where $\gamma_i$ is the corresponding degeneracy factor, $E_i^*$ the energy and $\mu_i^*$ the chemical potential of the quark.\\
All thermodynamical quantities, energy density $e$, entropy density $s$ as well as the
densities of the different particle species $\rho_i$, can be derived from the grand canonical potential. In our model it has the form:
\begin{equation}
	\frac{\Omega}{V}=-L_{int}-L_{meson}+\frac{\Omega_{th}}{V}-U
\end{equation}
Here $\Omega_{th}$ includes the heat bath of hadronic and quark quasi particles. The effective potential $U(\Phi,\Phi^*,T)$, which controls the dynamics of the Polyakov-loop, will be discussed in the following.
In our approach we adopt the ansatz proposed in \cite{Ratti:2005jh}:
\begin{eqnarray}
	U&=&-\frac12 a(T)\Phi\Phi^*\nonumber\\
	&+&b(T)ln[1-6\Phi\Phi^*+4(\Phi^3\Phi^{*3})-3(\Phi\Phi^*)^2]
\end{eqnarray}
 with $a(T)=a_0 T^4+a_1 T_0 T^3+a_2 T_0^2 T^2$, $b(T)=b_3 T_0^3 T$.\\

This choice of effective potential satisfies the $Z(3)$ center symmetry of the pure gauge Lagrangian. In the confined phase, $U$ has an absolute minimum at $\Phi=0$, while above the critical Temperature $T_0$ (for pure gauge $T_0 = 270$ MeV) its minimum is shifted to finite values of $\Phi$. The logarithmic term originates from the Haar measure of the group integration with respect to the SU(3) Polyakov loop matrix. The parameters $a_0, a_1, a_2$ and $b_3$ are fixed, as in \cite{Ratti:2005jh}, by demanding a first order phase transition in the pure gauge sector at $T_0=270$ MeV, and that the Stefan-Boltzmann limit of a gas of glouns is reached for $T \rightarrow \infty$.
Note that $T_0$ remains a free parameter to adjust the actual critical temperature,
of both phase transitions, when both, quarks and hadrons, couple to the scalar fields.\\

As has been mentioned above, the Lagrangian of the chiral model contains dilaton terms to model the scale anomaly. These terms constrain the chiral condensate, if the dilaton is frozen at its ground state value $\chi_0$. On the other hand, as deconfinement is realized, the expectation value of the chiral condensate should vanish at some point. On account of this we will couple the Polyakov loop to the dilaton in the following way:
\begin{equation}
	\chi=\chi_0 \ (1-0.5(\Phi\Phi^*))
\end{equation}
Assuming a hard part for the dilaton field which essentially stays unchanged and a soft part, which vanishes when deconfienemt is realized. 
Hence, allowing the chiral condensate to also approach zero.
\\

Until now all hadrons are still present in the deconfined and chirally restored phase.
Since we expect them to disappear, at least at some point above $T_c$, we have to include
a mechanism that effectively suppresses the hadronic degrees of freedom, when deconfinement is achieved.\\
In previous calculations baryons were suppressed by introducing a large baryon mass shift for
non-vanishing $\Phi$ \cite{Dexheimer:2009hi}.

In the following the suppression mechanism will be provided by excluded volume effects.
It is well known that hadrons are no point-like particles, but have a finite volume.
Including effects of finite-volume particles, in a thermodynamic model for hadronic matter, was proposed some time ago \cite{Hagedorn:1980kb,Baacke:1976jv,Gorenstein:1981fa,Hagedorn:1982qh}. We will use an ansatz similar to that used in \cite{Rischke:1991ke,Cleymans:1992jz}, but modify it to also treat the point like quark degrees of freedom consistently.\\

If one introduces a particle of radius $r$ into a gas of the same particles, then the volume excluded is not just the simple spherical volume, but one-half times the volume of a sphere with radius $2r$:
\begin{equation}
	v=\frac12 \cdot \frac43 \pi (2r)^3
\end{equation}
 It is easy to understand that if all other particles also have a radius $r$ then the excluded volume is much bigger than just the volume of a single particle.\\
We expect the volume of a meson to be smaller than that of a baryon. We introduce the quantity $v_{i}$ which is the volume excluded of a particle of species $i$ where we only distinguish between hadronic baryons, mesons and quarks. Consequently  $v_{i}$ can assume three values:
\begin{eqnarray}
 v_{Quark}&=&0 \nonumber \\
 v_{Baryon}&=&v \nonumber \\
 v_{Meson}&=&v/a \nonumber \\
\end{eqnarray}
where $a$ is a number larger than one. In our calculations we assumed it to be $a=8$, which would mean that the radius $r$ of a meson is half of the radius of a baryon.
Note that at this point we neglect any possible Lorentz contraction effects on the excluded volumes as introduced in \cite{Bugaev:2000wz,Bugaev:2008zz}.

The modified chemical potential $\widetilde{\mu}_i$, which is connected to the real chemical potential $\mu_i$ of the $i$-th particle species, is obtained by the following relation:
\begin{equation}
	\widetilde{\mu}_i=\mu_i-v_{i} \ P
\end{equation}
where $P$ is the sum over all partial pressures. All thermodynamic quantities can then be calculated with respect to the temperature $T$ and the new chemical potentials $\widetilde{\mu}_i$. To be thermodynamically consistent, all densities ($\widetilde{e_i}$, $\widetilde{\rho_i}$ and $\widetilde{s_i}$) have to be multiplied by a volume correction factor $f$, which is the ratio of the total volume $V$ and the reduced volume $V'$, not being occupied:
\begin{equation}
	f=\frac{V'}{V}=(1+\sum_{i}v_{i}\rho_i)^{-1}
\end{equation}

Then the actual densities are:
\begin{eqnarray}
e&=&\sum_i f \ \widetilde{e_i} \\
\rho_i&=&f \ \widetilde{\rho_i} \\
s&=&\sum_i f \ \widetilde{s_i}
\end{eqnarray}

Note that in this configuration the chemical potentials of the hadrons are decreased by the quarks, but not vice versa. In other words as the quarks start appearing they effectively suppress the hadrons by changing their chemical potential, while the quarks are only affected through the volume correction factor $f$.\\
Our implementation of finite-volume corrections
as outlined above is a simple approach with as few parameters as possible
and can be improved upon in various ways. For one, hadrons differ in size.
The size of a hadron could even be density or temperature dependent \cite{Kapusta:1982qd}. In addition, the excluded-volume parameter
of a particle does also depend on the density of the system (at dense packing a particle excludes
effectively less volume). Furthermore there might be dynamic effects due to interaction which increase with temperature and/or density. Such effects could be introduced to some extent, making the model even more depend on parameters. Such a finetuning could help to better describe e.g. lattice data but requires extensive work on the model. At least some aspects of a density dependend excluded volume can be considered in future works. 
However, in the following, one should regard the variables $v$ and $a$ as effective parameters
for capturing the qualitative effect of an excluded volume correction, which suppresses the hadrons in the quark phase.\\
We would like to stress that these volume corrections enable us to describe a phase transition from hadronic
to quark degrees of freedom, having only one single partition function for both phases, in a thermodynamic consistent
manner. Furthermore the volume corrections we apply are physically well motivated and are thoroughly discussed in
older and recent literature \cite{Hagedorn:1980kb,Baacke:1976jv,Gorenstein:1981fa,Hagedorn:1982qh,Rischke:1991ke,Cleymans:1992jz,Bugaev:2000wz,Bugaev:2008zz,Satarov:2009zx}. They model the fact that hadrons generate a repulsive hard-core interaction for the 
other particles in the system. This is not necessarily related to confinement. Therefore, a volume correction from 
the remaining mesons beyond $T_c$ that also affect the quarks is not a contradiction to the fact that quarks can 
propagate freely, it is part of the residual interaction in the system, quarks feel repulsion interacting with 
the mesons. Our description of the excluded volume effects is at this stage simplified and parameter dependent 
(yet thermodynamically consistent).\\


\section{Results at vanishing net baryon density}

In this section we concentrate on the properties of the model at vanishing chemical potential. Here lattice calculations suggest a crossover from the hadronic to the quark phase. Different lattice groups obtain different results for the phase transition temperature ranging from $T_c= 160$ MeV to $200$ MeV \cite{Aoki:2009sc,Detar:2007as}.
For all following results we set $T_0$, the free parameter of the Polyakov-potential, to $T_0=235$ MeV and the excluded volume parameter $v= 1 {\rm fm}^3$. This leads to a critical temperature of $T_c \approx 183$ MeV ($T_c$ is defined as the temperature with the largest change in the order parameters as a function of the temperature). In this section we will also distinguish results obtained when the Polyakov loop is coupled to the dilaton in the above described manner (solid lines), and those where the dilaton is not coupled to the Polyakov loop (dashed lines).\\
The lattice data referred to in the following sections are taken from the HotQCD collaboration \cite{Bazavov:2009zn}. Here different actions (p4, asqtad) and lattice spacings ($N_{\tau}= 6,8$) were compared. Note that the transition region extracted from the lattice data lies between $185$ and $195$ MeV.
The reader should keep in mind that different lattice groups get significantly differing results on all observables. 
This indicates that the systematic uncertainties on lattice data are still very large 
(much larger than the statistical errors which are usually plotted). In fact, recent lattice results of the HotQCD group \cite{Bazavov:2010sb} with new actions in order to improve the description of hadrons on the lattice, point to the importance of hadrons to describe the phase transition as well as explicitly state the rather slow and smooth transition from confinement 
to deconfinement with a wide intermediate region.
There are even attempts to combine results on thermodynamics from the lattice with those from 
a hadronic resonance gas which is expected to be the correct description of matter below $T_c$ \cite{Huovinen:2009yb}.
We therefore do not expect to get a good agreement of our results on thermodynamics with the lattice data below $T_c$.
\\
\begin{figure}[t]
\includegraphics[width=0.5 \textwidth]{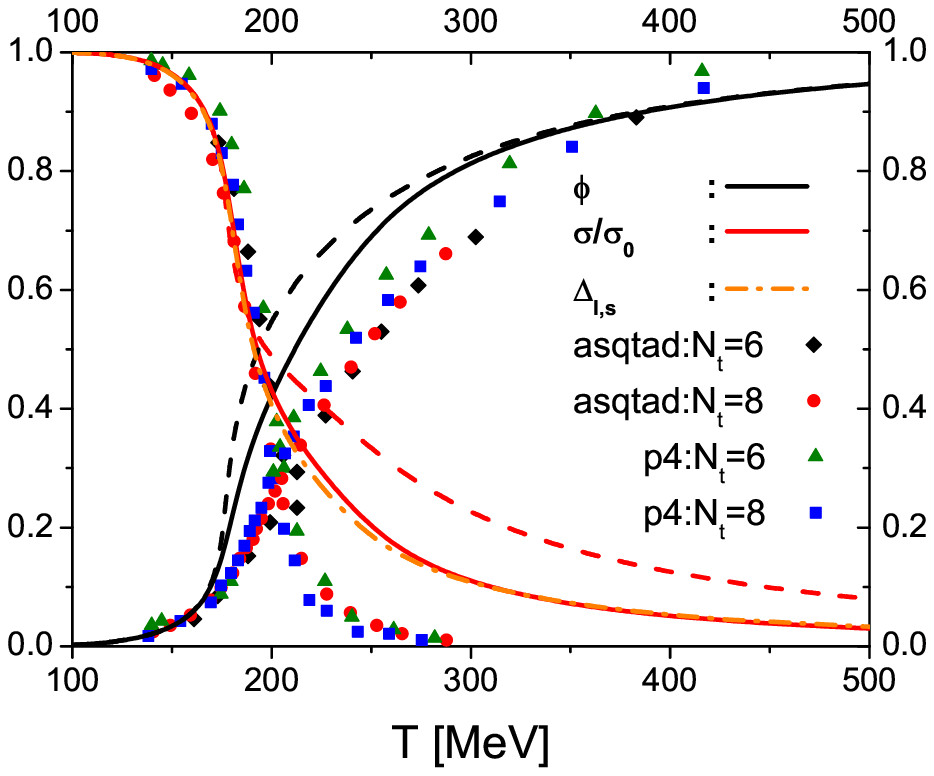}
\includegraphics[width=0.5 \textwidth]{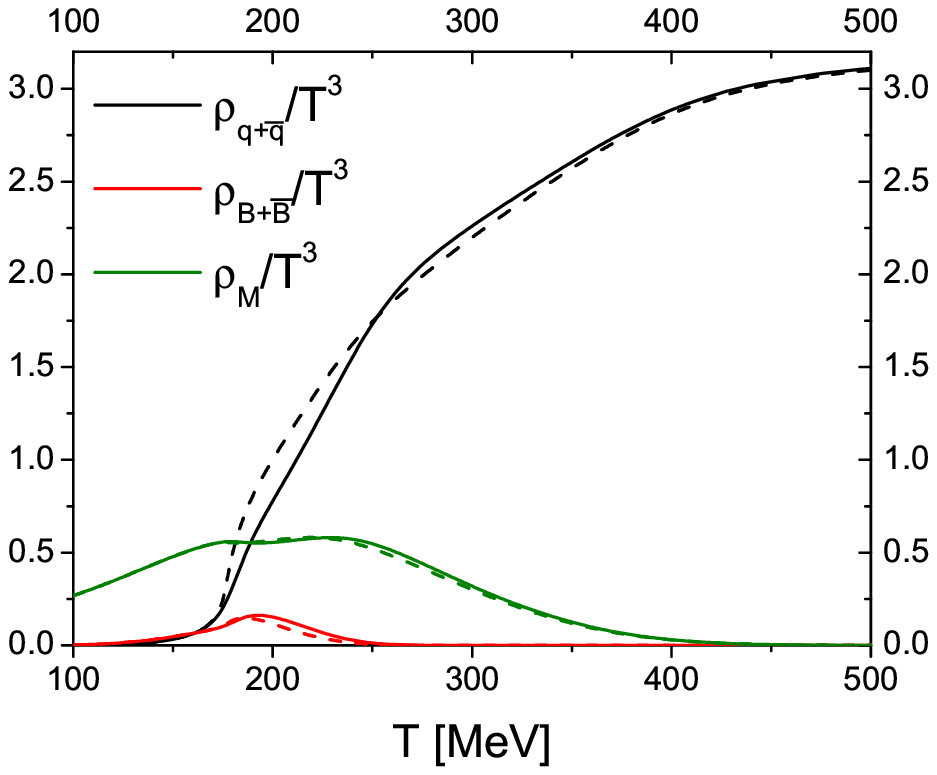}
\caption{\label{mu0sig}\bfseries{Left}\mdseries : The normalized order parameters for the chiral (red lines), and deconfinement (black line) phase transition as a function of $T$ at $\mu_B = 0$. Also indicated is the subtracted chiral condensate as defined in the text (orange dash-dotted line). The dashed lines depict results without the dilaton-Polyakov loop coupling. The symbols denote lattice data for the subtracted chiral condensate from \cite{Bazavov:2009zn}, using different lattice actions (asqdat and p4) and lattice spacings $N_{\tau}$.\\
\bfseries Right\mdseries : Total particle number densities for the different particle species devided by $T^3$ as a function of $T$ at $\mu_B = 0$. The black line shows the total number of quarks+antiquarks per volume while the green (dotted) line refers to the total meson density and the red (dashed) line to the number density of hadronic baryons+antibaryons. The dashed lines depict results without the dilaton-Polyakov loop coupling.}
\end{figure}

Figure (\ref{mu0sig} left) shows the temperature dependence for the order parameters of both,
the deconfinement ($\Phi$), and chiral ($\sigma$) phase transition, extracted from our model and compared to lattice data. Both order parameters change smoothly with temperature. The critical temperature is found to be equal for both
phase transitions. 
The lattice results represent a quantity which is called the subtracted chiral condensate ($\Delta_{l,s}$) and which is defined in the following way:
\begin{equation}
	\Delta_{l,s}(T)=\frac{\sigma(T)-m_q/m_s \zeta(T)}{\sigma(0)-m_q/m_s \zeta(0)}
\end{equation}
Here $m_q$ and $m_s$ refer to the bare mass of up, down and strange quarks.\\
Note that the value of the chiral condensate $\sigma$ approaches zero only slowly. 
	This originates from the dilaton contribution to the scalar potential in this model
	(the logarithmic term in eqn.(3) prevents the value of sigma to drop as fast as expected), which 
	generates a repulsive term for small values of $\sigma$. Therefore the temperature dependence 
	of the chiral condensate compares less favorably to lattice results than PNJL type models.
	A simple ansatz to solve this problem would be to simply remove the dilaton contribution from the model,
	but in the current parametrization of the hadronic model the dilaton contribution is essential for the correct 
	description of the ground state of nuclear matter.
	To resolve this problem we would have to introduce a different coupling of the dilaton to the hadrons (and quarks)
	 and fields (as for example outlined in \cite{Ellis:1991qx}, here part of the baryon mass is generated through coupling to the dilaton and
	part through coupling to the chiral condensate). This way one may achieve a satisfactory description of nuclear
	ground state and a satisfactory behavior of the chiral condensate above $T_c$. Such a major change of the
	chiral model is out of the scope of the current work and will be pursued in the future.
	\\
Like in the PNJL model the parameters of the Polyakov potential are fixed by a fit to pure glue lattice data.
Hence, the value of the Polyakov Loop increases somewhat faster as a function of temperature than in recent 
lattice calculations including quarks. The same behavior can be observed, when a PNJL type of model \cite{Schaefer:2009ui} is 
compared against the newest lattice results as a function of temperature (and not $T/T_c$ which can be misleading at some point).

\begin{figure}[t]
\includegraphics[width=0.5 \textwidth]{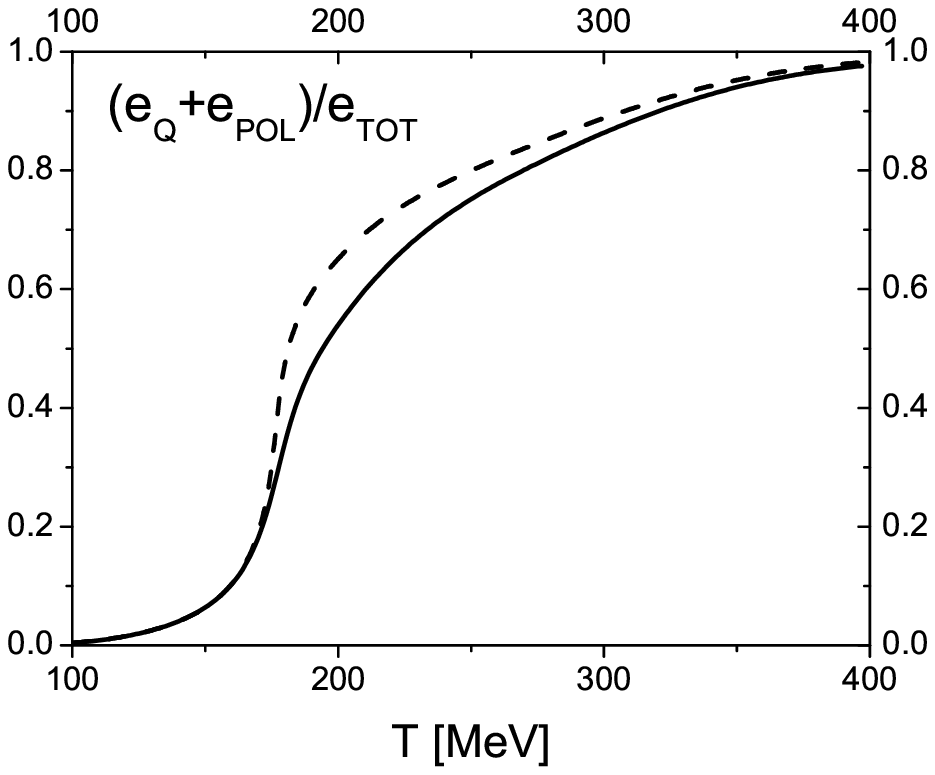}
\includegraphics[width=0.5 \textwidth]{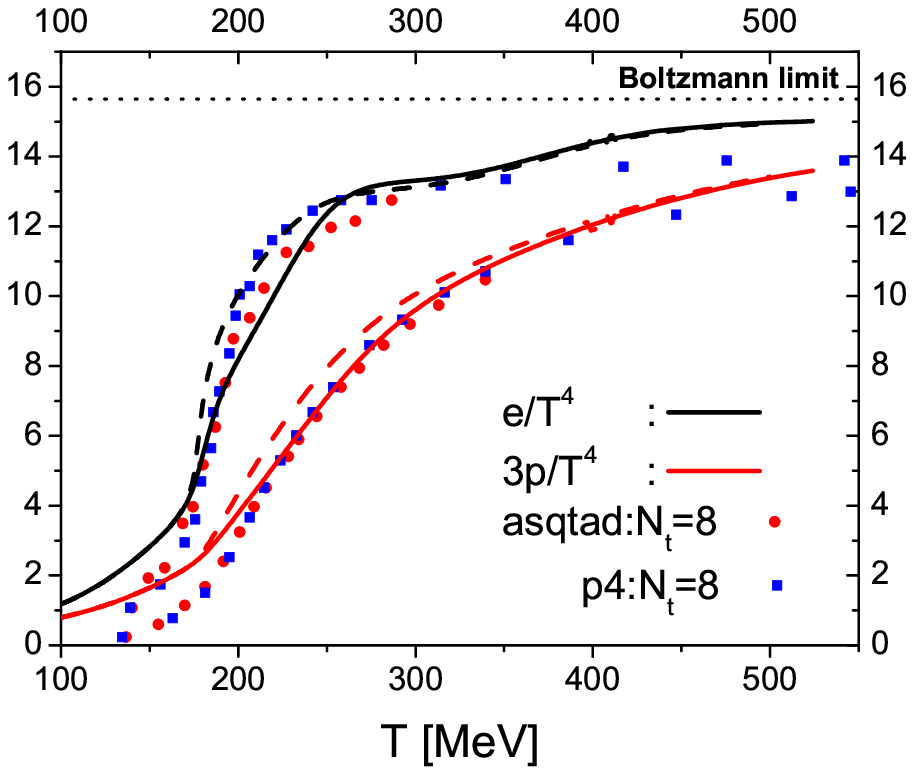}
\caption{\label{mu0mix}
\bfseries Left \mdseries: The fraction of the total energy density that can be assigned to the quark-gluon phase ($e_{QGP}$ contains the energy of the quarks and the Polyakov potential) as a function of $T$ at $\mu_B = 0$. The dashed lines depict results without the dilaton-Polyakov loop coupling.\\
\bfseries Right \mdseries: Three times the pressure (red lines) and energy density (black lines) over $T^4$ as a function of $T$ at $\mu_B = 0$. The green dotted line indicates the Boltzmann limit for an ideal gas of three massless quarks and gluons. The symbols denote lattice data from \cite{Bazavov:2009zn}, using different lattice actions (asqdat and p4) and lattice spacings $N_{\tau}$. The dashed lines depict results without the dilaton-Polyakov loop coupling.}
\end{figure}


Figure  (\ref{mu0sig} right) shows the total densities of quarks plus antiquarks (black lines), mesons (green lines) and baryons plus antibaryons (red lines). Below the critical temperature hadrons are the dominant degree of freedom. When the quark number increases around $T_c$, they begin to suppress the hadrons. It is remarkable that the hadrons are still present, and not negligible, up to about $2.0 \ T_c$ \cite{Stoecker:1980uk}. Especially the mesons contribute strongly to all thermodynamic quantities, since they are quite less suppressed than the baryons ($v_M < v_B$).

Above $2 \ T_c$ the hadrons are effectively squeezed out of the system by the presence of the quarks. 
To emphasize this change in degrees of freedom, Figure(\ref{mu0mix} left) shows the fraction of the total energy density which stems from the quarks and gluons (more precisely the Polyakov potential). As expected for a crossover both degrees of freedom (hadrons and quarks) are present in the temperature range from $0.75-2 \ T_c$. However, around $T_c$ the fraction of the energy density, due to quarks and gluons increases rapidly. It converges to unity at around 2 times $T_c$.

Let us now take a closer look at different thermodynamic quantities. Figure (\ref{mu0mix} right) displays the energy density (black curve) and three times the pressure (red dashed curve), both over $T^4$ compared to lattice data \cite{Bazavov:2009zn}. In the limit of infinite temperature, both quantities should converge to the Stefan-Boltzmann limit of an ideal gas of quarks and gluons. This limit is indicated as a green dashed line. The strong increase in energy density around $T_c$ reflects the rapid change of the relevant degrees of freedom. At three times the critical temperature the energy density is slowly converging to the Stefan-Boltzmann limit, while the pressure is converging even slower as it was also observed in PNJL calculations \cite{Roessner:2006xn}. At temperatures below $T_c$ our calculation gives larger values for the pressure and the energy density.

Around $1.5 \ T_c$ one can observe a slight 'dip' in the energy density. This 'dip' is connected to the correction factor $f$ of the excluded volume corrections. As has been shown above, the hadronic contribution to the densities disappears only at two time $T_c$ and therefore they still exclude some portion of the volume for the quarks. The 'dip' therefore indicates the disappearance of volume correction factors for the quark phase.

In the high temperature limit, where only the quarks (and gluons) remain in the system the energy density and pressure both slightly exceed the data from lattice calculations.\\


\begin{figure}[t]
\includegraphics[width=0.5 \textwidth]{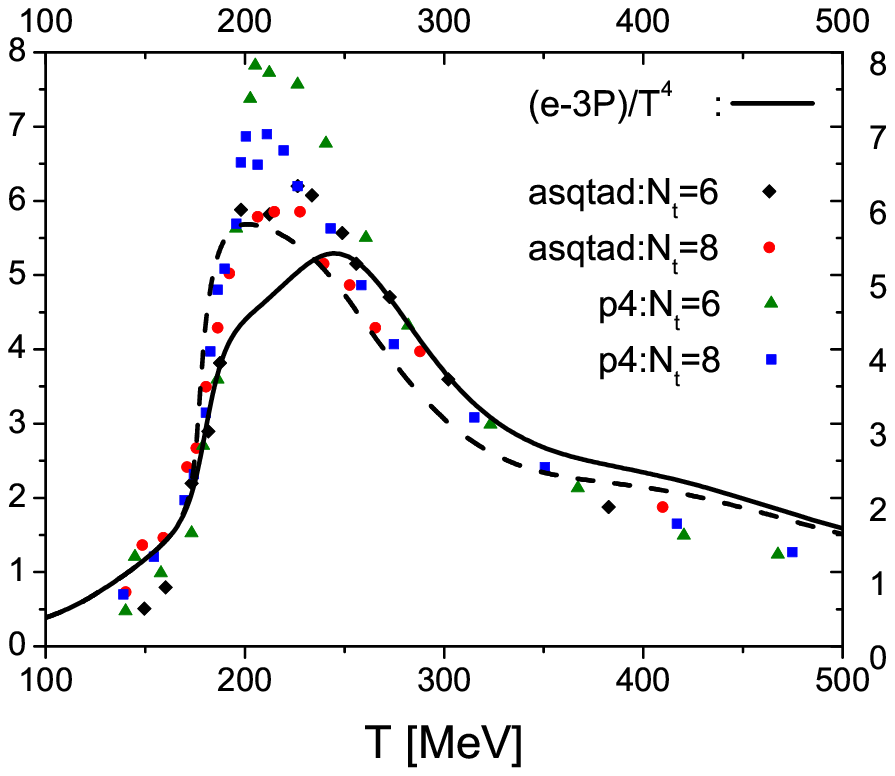}
\includegraphics[width=0.5 \textwidth]{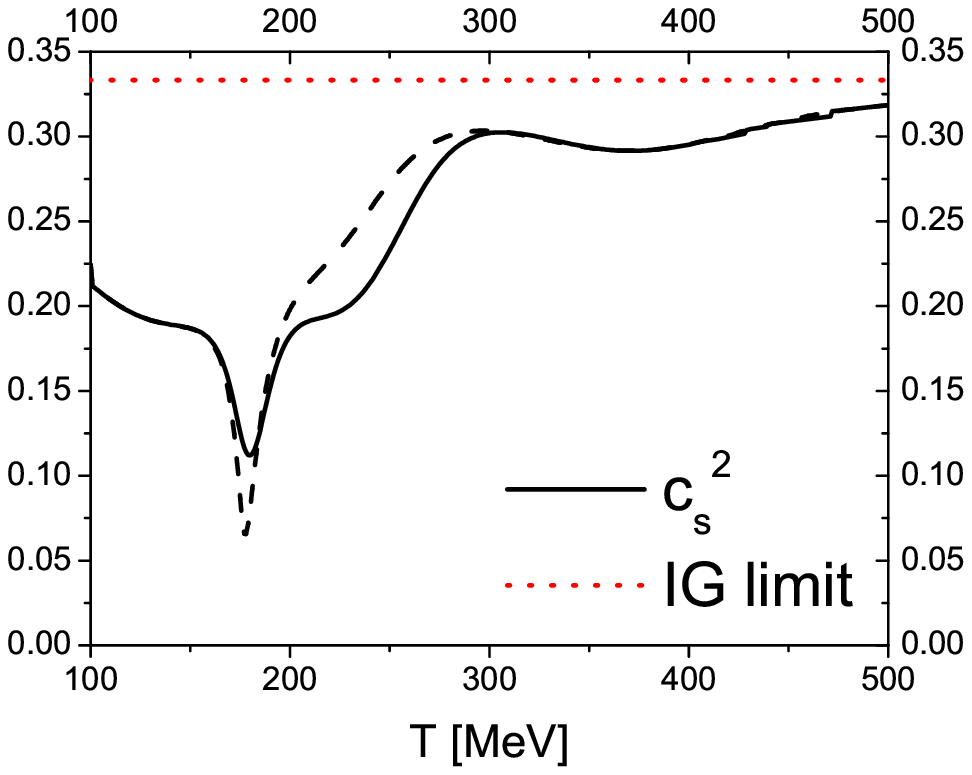}
\caption{\label{mu0e3p} \bfseries Left \mdseries: Energy density minus three times the Pressure over $T^4$ as a function of $T$ at $\mu_B = 0$. Also referred to as the interaction measure. The symbols denote lattice data from \cite{Bazavov:2009zn}, using different lattice actions (asqdat and p4) and lattice spacings $N_{\tau}$. The dashed lines depict results without the dilaton-Polyakov loop coupling.\\
\bfseries Right \mdseries: The speed of sound squared, as a function of $T/T_c$ at $\mu_B = 0$. $1/3$ is the ideal gas limit. The dashed lines depict results without the dilaton-Polyakov loop coupling.}
\end{figure}

Figure (\ref{mu0e3p} left) displays the difference of the energy density and three times the pressure over $T^4$ (black solid line). This quantity is also referred to as the 'interaction measure' in lattice calculations. In the
Stefan-Boltzmann limit it is $0$, while it shows a peak slightly above $T_c$. The height of the peak in our model is comparable to the lower bound from lattice studies \cite{Bazavov:2009zn}, while its value at large $T$ is a little bit above that from lattice calculations, because chiral restoration is not fully achieved in our model.\\
We want to point out that, even though our model results below $T_c$ agree well with the lattice results for the interaction measure, the corresponding values for the energy density and pressure differ considerably, when we compared to lattice data (Figure \ref{mu0mix} right). This points out that the newest lattice results still do not have the resolution to describe the hadronic part of the heat bath sufficiently well \cite{Bazavov:2009zn}. Thus, interaction effects of hadronic states in the hot system are most likely not correctly taken into account in the current lattice data.      

Note that our model gives a much better description of the interaction measure as it does for the order parameters when compared to lattice. This might indicate that the conversion from the behavior of the Polyakov loop to the bahavior of thermodynamic quantities might not be as proposed in a PNJL type approach.

An important property of a hot and dense nuclear medium is the speed of sound ($c_s$):
\begin{equation}
c_s^2 = \left. \frac{d p}{d e}\right|_{\mu=0}
\end{equation}
It is not only closely related to expansion dynamics but also controls the way perturbations (sound- and shock-waves) travel
through the fireball \cite{Stoecker:1980vf}. Figure (\ref{mu0e3p} right) shows the square of the speed of sound as a function of temperature. As the temperature increases towards $T_c$ one can clearly observe a softening of the EoS due to the crossover. At very high temperature the speed of sound again converges toward its ideal gas limit of $c_s^2 \rightarrow 1/3$. The dip above $T_c$ is again related to the excluded volume corrections. Note that even though the change of degrees of freedom from hadrons to quarks proceeds as a crossover, there is still a substantial softening (i.e. $c_s^2$ goes down to $0.07$). This behavior is comparable to results obtained with different versions of the PNJL model \cite{Schaefer:2009ui,Rossner:2007ik}. It is a result of the fit of the Polyakov loop potential to pure glue data, resulting in a steeper increase of the order parameter when compared to lattice results including quarks.\\


Overall, the results are not very sensitive on the exact values used in the excluded-volume part of the model.
Varying the volume parameter $v$ by a factor of $2$ does not alter the temperature dependence of 
the Polyakov loop as it is controlled mainly by the Polyakov potential (and therefore the parameter $T_0$). Even the phase transition temperature $T_c$ of the chiral phase transition is not affected by the volume parameter. For smaller values of the volume the chiral phase transition becomes slightly steeper (a faster increase with temperature). The thermodynamic quantities change at maximum about $10\%$ around $T_c$, but the qualitative behavior of the hadrons beeing suppressed by the quarks stays unchanged. It is rather a question of how much the hadrons get suppressed at a given temperature and therefore
how much they still contribute in the region of phase coexistence. At some point above $T_c$ the hadrons are removed for any value of the volume parameter.
One could probably do some fine tuning to whatever digit desired to get a maybe slightly better description 
of the lattice data, but that was not the intent of the manuscript. The value of $1 \ \rm{fm}^3$ for the
volume parameter is a rough estimate of a baryons volume which was widely used in the literature.

\section{Results at zero temperature}

\begin{figure}[t]
\includegraphics[width=0.5 \textwidth]{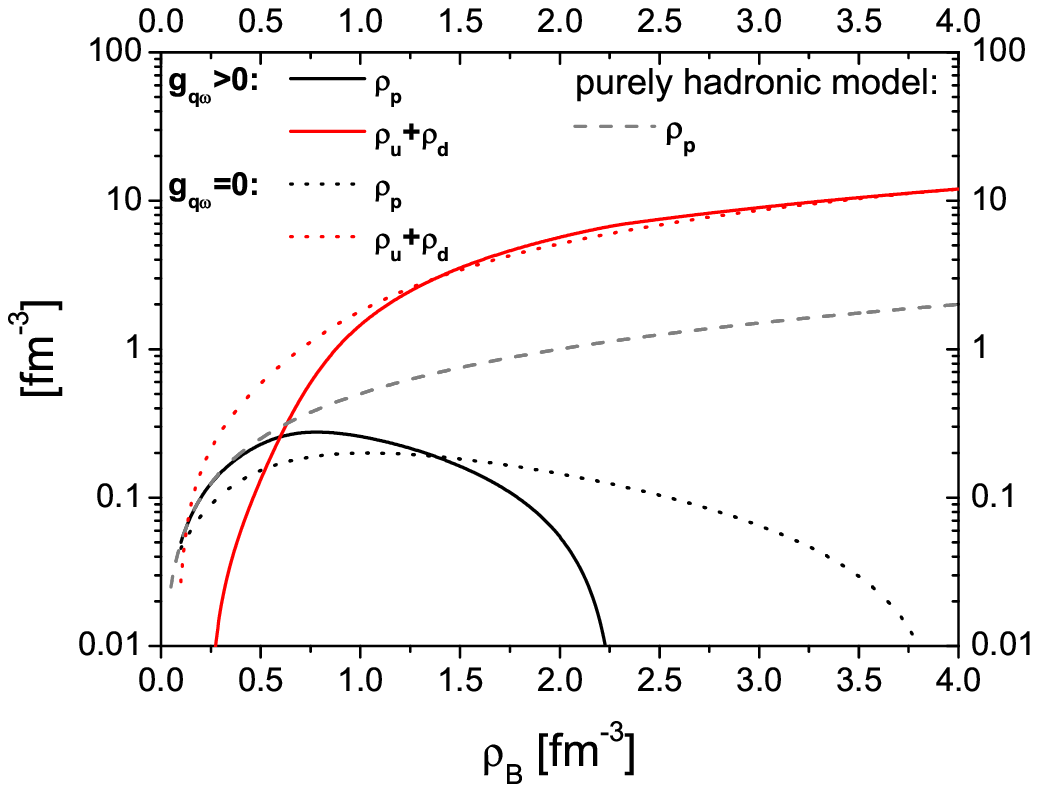}
\includegraphics[width=0.5 \textwidth]{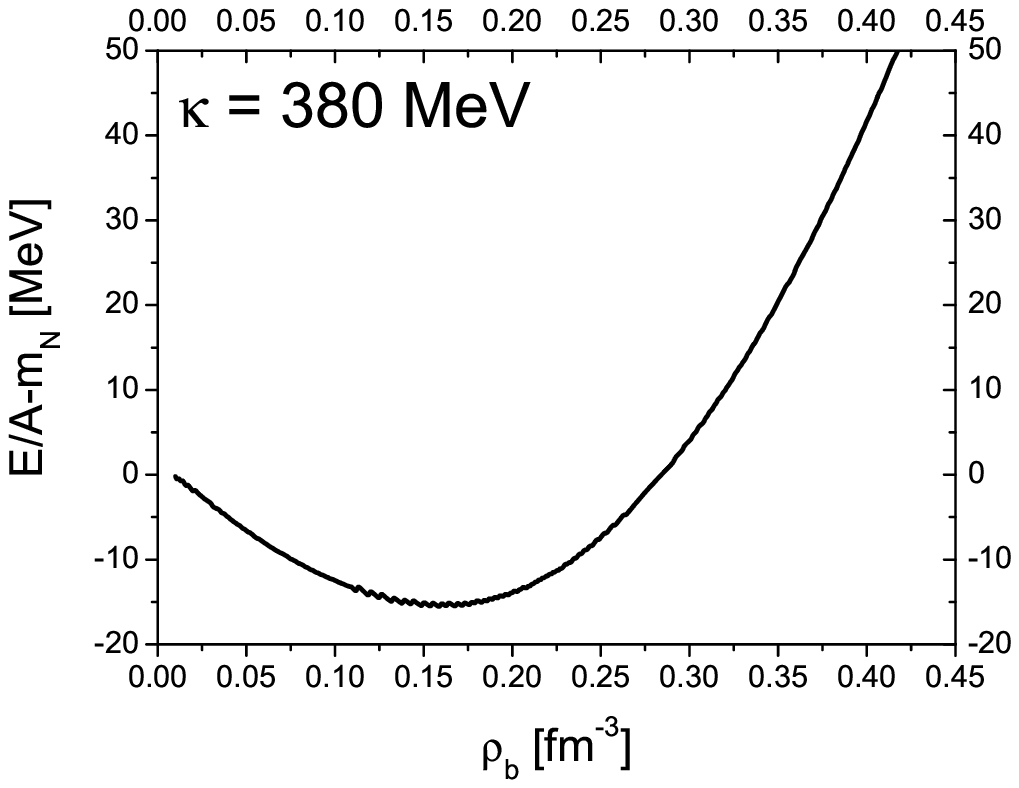}
\caption{\label{t0dens} \bfseries Left \mdseries: Densities of the different particle species a a function of net baryon density at $T=0$. The grey dashed line is the result from the purely hadronic model and serves as a baseline comparison. The red lines show the results for the quark densities while the proton densities are depicted in black. Results with a quark vector interaction strength of $g_{q \omega} = 3.0$ are shown as solid lines while the dashed lines are results with $g_{q \omega} = 0$\\
\bfseries Right \mdseries: Binding energy per nucleon as a function of net baryon density at $T=0$ for $g_{q \omega} = 3.0$. The minimum is located at $\approx 0.16 fm^{-3}$ and the corresponding binding energy is $E/A -m_N \approx -16 MeV$. The resulting incompressibility is $\kappa \approx 380 MeV$}
\end{figure}

Most of the physical observables we have accurate knowledge of are measured in the vacuum and, concerning the scales used in this work, at almost zero temperature. Important observables for our work are for example the vacuum masses of the hadrons and the properties of nuclear ground state matter. So before extending our work to the realm of finite baryon density and finite temperature we need to make sure that our model gives reasonable results at zero temperature. First we want to investigate the behavior of the different particle densities, at $T=0$, as a function of net baryon density. Here the repulsive vector interaction, transmitted by the vector field $\omega$, starts to play a more important role than the attractive interaction originating from the $\sigma$ field. In our model the vector interaction strength $g_{n\omega}$ of the nucleons can easily be constrained by demanding reasonable values for the nuclear binding energy and saturation density. This is not the case for the quark vector-interaction strength. The only reasonable constraint on their part would be to demand that there are no free quarks present in the nuclear ground state.\\
Figure (\ref{t0dens} left) shows the densities of quarks (red lines) and protons (black lines) as a function of net baryon density compared to the nucleon density from the purely hadronic model (grey dashed line). In the case of no repulsive quark vector interaction (dotted lines) the free quarks appear already before the ground state density. This would mean that there is no nuclear liquid-gas phase transition and no physical nuclear ground state. If we introduce a finite quark vector interaction strength of $g_{q \omega} = 3.0 \approx g_{n\omega}/3 $ the quarks appear only at larger densities.\\

In this scenario we can obtain reasonable values for the nuclear ground state saturation density ($\rho_0 \approx 0.16 \ \rm{fm}^{-3}$) as well as the binding energy ($E/A -m_N \approx -16$ MeV) as can be seen in Figure (\ref{t0dens} right). We can also calculate the incompressibility modulus at ground state density, defined as $\kappa=9 (dp/d\rho_b)_{T=0,\rho_b=\rho_0}$ and obtain a result of $\kappa \approx 380 MeV$. This value is somewhat larger as expected \cite{Blaizot:1980tw} and is due to the hard core repulsive interactions which make the system rather incompressible. We have checked that in fact the speed of sound does stay below $c_s = 1$ for any density or temperature.

\section{Results at finite net baryon density and temperature}

\begin{figure}[t]
\includegraphics[width=0.5 \textwidth]{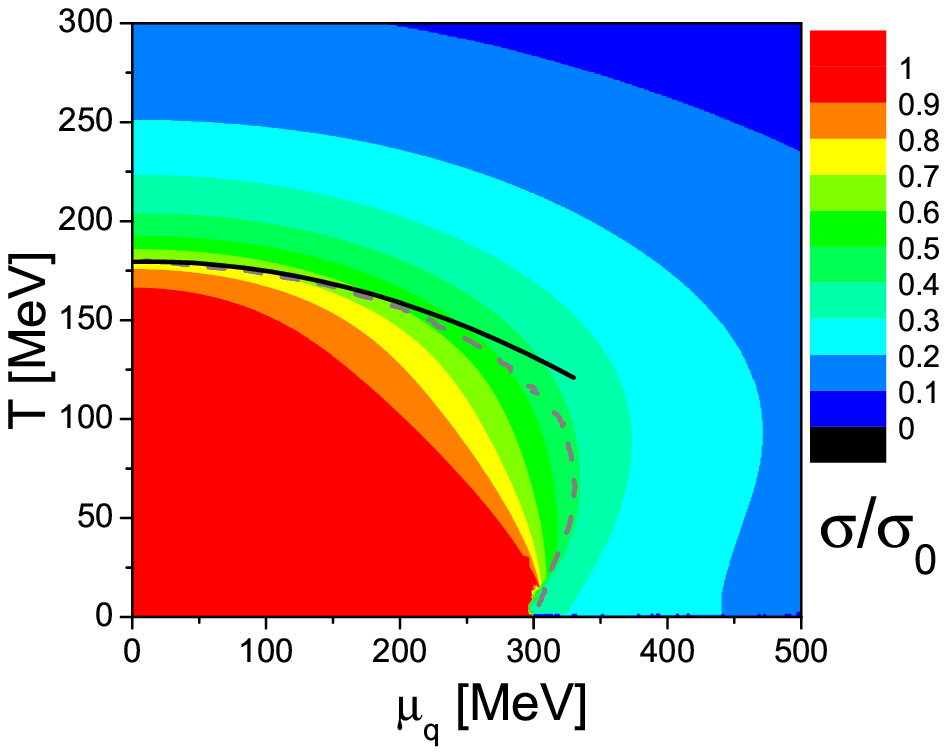}
\includegraphics[width=0.5 \textwidth]{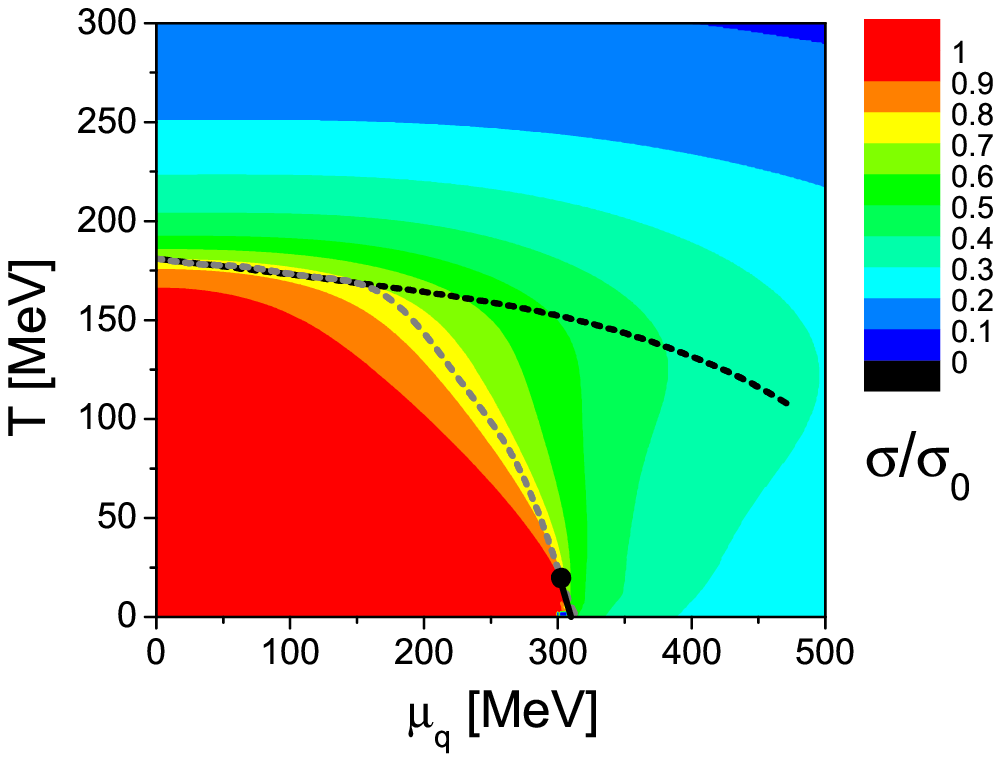}
\caption{\label{tmusigno} \bfseries Left \mdseries: Contour plot of the chiral condensate devided by its ground state value as a function of temperature and quark chemical potential. The value for the quark repulsive interaction is $g_{q \omega} = 0$. The dashed grey line indicates where the change of the chiral condensate with respect to $T$ and $\mu_q$ has a maximum. The solid black line shows the same for the change of the Polyakov loop.\\
 \bfseries Right \mdseries: Contour plot of the chiral condensate divided by its ground state value as a function of temperature and quark chemical potential. The value for the quark repulsive interaction is $g_{q \omega} = 3$. The dashed grey line indicates where the change of the chiral condensate with respect to $T$ and $\mu_q$ has a maximum. The dashed black line shows the same for the change of the Polyakov loop. The liquid gas phase transition, which is first order at $T=0$ is shown as the black solid line with a critical endpoint at $T_{cep} \approx 16 MeV$.}
\end{figure}


Since our model does not sustain the difficulties that lattice calculations have, when going to finite densities, we can simply extend our investigations to finite chemical potentials ($\mu_B=  3 \mu_q \neq 0$) and temperature. As for the case at $T=0$ any repulsive vector interaction for the quarks may change the picture of the phase diagram. The value of the repulsive interaction strength $g_{q \omega}$ is not constrained by any first principle-calculation. It could even be a function of temperature and density. For simplicity we will compare two cases, $g_{q \omega}=0$ and $g_{q \omega}=3$, to show the qualitative changes of the phase diagram when different values of the vector coupling strength are assumed.\\
For calculations at finite $\mu_B$, the strange quark chemical potential $\mu_s$ also plays an important role. If one assumes, that the total net strangeness is globally conserved, hadronic chemistry induces a non zero chemical potential for the strange quark, while it vanishes in the case of a free quark gas 
. In the present work we always constrain the net strangeness to be zero, but one should also investigate the phase structure of a system where this is not fulfilled.
Work along this line is in progress.\\

\begin{figure}[t]
\includegraphics[width=0.5 \textwidth]{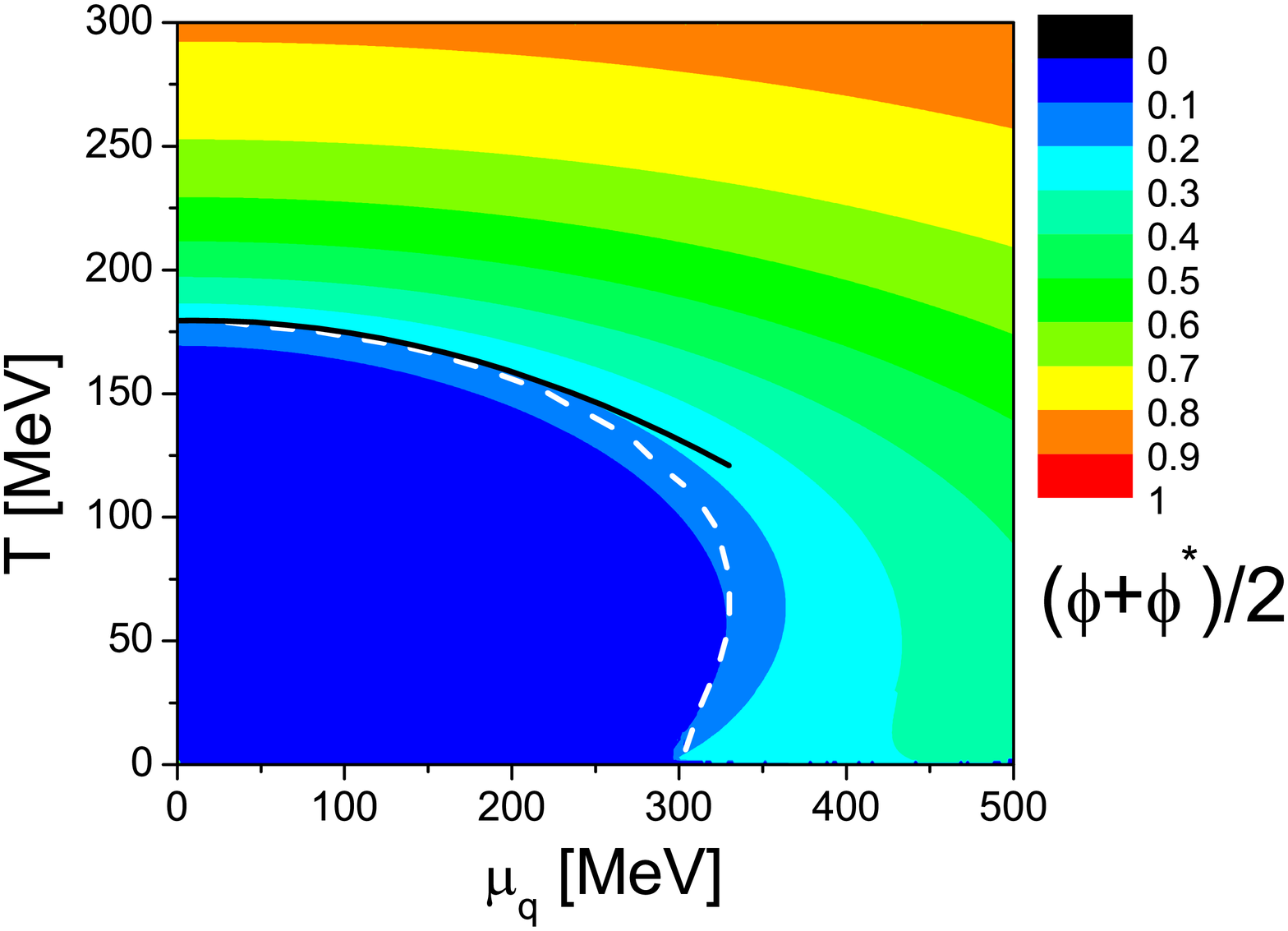}
\includegraphics[width=0.5 \textwidth]{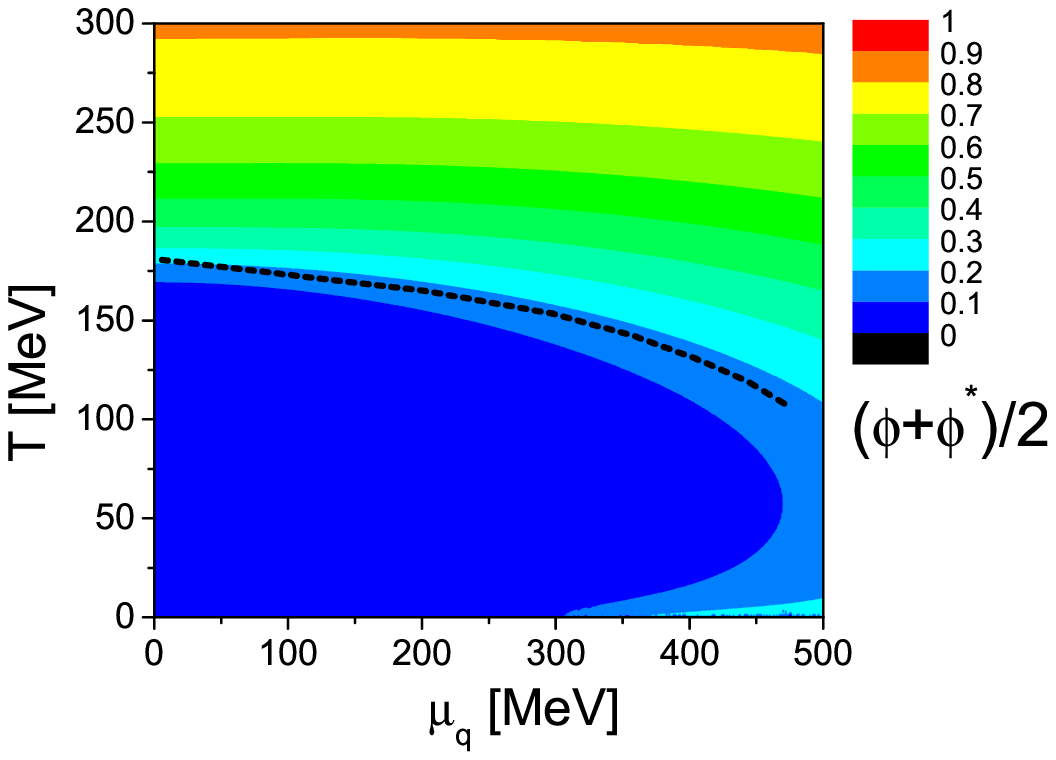}
\caption{\label{tmupolno}\bfseries Left \mdseries: Contour plot of the Polyakov loop as a function of temperature and quark chemical potential. The value for the quark repulsive interaction is $g_{q \omega} = 0$. The dashed grey line indicates where the change of the chiral condensate with respect to $T$ and $\mu_q$ has a maximum. The solid black line shows the same for the change of the Polyakov loop.\\
\bfseries Right \mdseries: Contour plot of the Polyakov loop as a function of temperature and quark chemical potential. The value for the quark repulsive interaction is $g_{q \omega} = 3$. The dashed black line indicates where the change of the Polyakov loop with respect to $T$ and $\mu_q$ has a maximum.}
\end{figure}


Figures (\ref{tmusigno} left) and (\ref{tmusigno} right) display a contour plot of the chiral condensate $\sigma$ (normalized to its ground state value $\sigma_0$) in the $T-\mu_q$ phase diagram for the two values of the repulsive quark interaction strength. It is apparent to see that in both cases the chiral phase transition is
a smooth crossover for all chemical potentials. Only at very low temperatures the iso-$\sigma$ lines converge to a first order phase transition. In the case of $g_{q \omega}=3$ this can be identified with the liquid-gas phase transition, which is first order at zero temperature (displayed in the plot as the solid black line with a critical endpoint at $T_c \approx 16 MeV$). As mentioned above, if the repulse vector interaction strength of the quarks is smaller, they already appear at or before the nuclear ground state and cause the first order jump in the order parameter. At high temperature and high chemical potentials the lines of constant $\sigma$ are even further apart than at vanishing net baryon density, indicating an even smoother crossover than at $\mu_q = 0$. We can still calculate the derivative of the order parameter with respect to the temperature and chemical potential. The grey dashed lines indicate where this gradient, with respect to the temperature and chemical potential, has his maximum in the phase diagram. The black dashed lines indicate a second, very small maximum in this derivative, which is caused be a rapid change in the deconfinement order parameter, the Polyakov loop. This suggests that the largest change in the chiral condensate is governed by hadronic interactions and not by the appearance of the quarks.

The hadrons dominate the change in the chiral condensate because they dominate the scalar baryon density in the region where 
the change in the condensate appears. At low temperatures the quarks are suppressed due to the Polyakov potential and at intermediate temperature there are simply much more hadronic degrees of freedom that couple to the chiral fields as there are quarks degrees of freedom.
If one would introduce hadrons as bound states of quarks then of course the change would be dominated by those bound states 
(again because there are just many more colour neutral states possible then quark states). But then it would again be the 
hadrons that dominate the chiral dynamics and not the free quarks.

\begin{figure}[t]
\centering
\includegraphics[width=0.7 \textwidth]{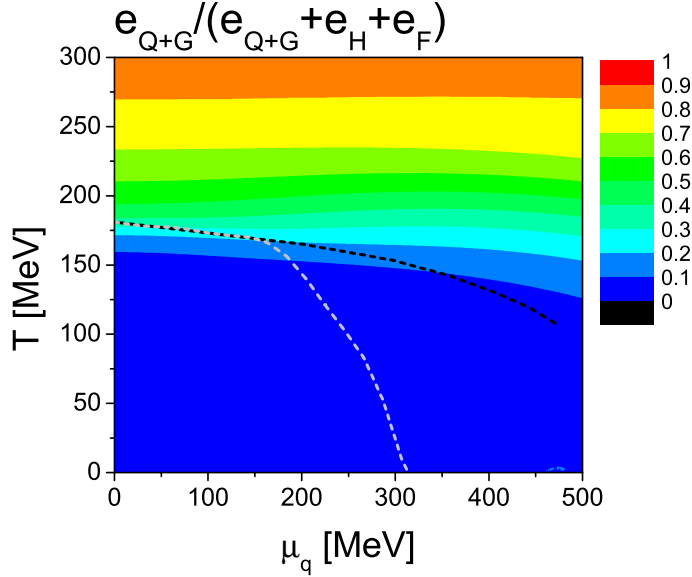}
\caption{\label{tmumix}Contour plot of the fraction of the total energy density which comes from the quark contribution and the Polyakov loop potential. The value for the quark repulsive interaction is $g_{q \omega} = 3$. The dashed grey line indicates where the change of the chiral condensate with respect to $T$ and $\mu_q$ has a maximum while the solid black line shows the same for the change of the Polyakov loop. $e_F$ symbolizes the energy density of the mean fields.}
\end{figure}

The observation of a smooth crossover can also be found in the contour plot of the Polyakov loop as a function of $T$ and $\mu_q$, in Figures (\ref{tmupolno} left) and (\ref{tmupolno} right). Here again results for two different values of the repulsive vector interaction are shown. The black dashed lines indicate the maximum of the derivative of the Polyakov loop with respect to temperature and chemical potential. There is a visible difference in the chiral (grey dashed) and deconfinement (black dashed) phase transitions at high chemical potentials. While the chiral condensate already drops at the liquid gas phase transition $\mu_q \approx 313$ MeV, the Polyakov loop remains small until much higher chemical potentials.

This behavior is very interesting as it means that, in some parts of the phase diagram, chiral symmetry is partially restored while quarks are still confined.
This can be made more clear if one looks at the fraction of the total energy density which can be contributed to quarks and gluons (more precise quarks and the Polyakov potential).
This fraction ($e_{\rm QGP}/e_{\rm TOT}$) is plotted in Figure (\ref{tmumix}), using $g_{q \omega} = 3$, in the phase diagram where we again indicated the largest change in the chiral condensate (grey dashed line) and the Polyakov loop (black dashed line). Again one observes, at high chemical potentials and intermediate temperatures, a phase which consists quarks and hadrons, while chiral symmetry is partially restored.

The transition which is often called the 'quarkyonic phase transition' would correspond to the usual nuclear liquid gas phase 
transition in our model. The dominant degrees of freedom are the hadrons, dominated by very light baryonic states. 
There where attempts to modify the Polyakov potential in a way to make it depend explicitly on the chemical potential \cite{Dexheimer:2009hi}. This 
way one can strongly couple the chiral phase transition and the change in the Polyakov loop. This approach however would
introduce a contribution to the net baryon number from the Polyakov potential as well as a contribution to
the quark number susceptibility, nothing of which is seen in lattice calculations.

\section{Conclusion}

We presented an approach for modeling an EoS that respects the symmetries underlying QCD,
and includes the correct asymptotic degrees of freedom, i.e. quarks and gluons at high temperature and hadrons in the low-temperature limit. We achieve this by including quarks degrees of freedom and the thermal
contribution of the Polyakov loop in a hadronic chiral sigma-omega model. The hadrons are suppressed at high
densities by excluded volume corrections. Nevertheless, we observe a substantial hadronic contribution to the EoS up to temperatures of 2 times $T_c$.\\
We can show that the properties of the EoS are in qualitative agreement with lattice data at $\mu_B = 0$. Various quantities, like the pressure and energy density, are in good agreement with lattice data. Deviations from lattice results can be explained by the hadronic contributions and volume corrections.
In spite of a continuous phase transition, we obtain a considerably smaller value for the speed of sound around $T_c$ ($c_s^2 \approx 0.072$) when compared to lattice calculations \cite{Bazavov:2009zn}.
At finite baryon density, the transition from deconfined to confined matter proceeds as a smooth crossover for all values of $\mu_B$. The same is true for the chiral phase transition (except the liquid gas phase transition, which is of first order at very low temperatures). At high chemical potentials and low temperatures we find a very interesting phase structure. In this region chiral symmetry is partially restored, while deconfinement is not yet realized, thus creating an exotic form of matter.

\section*{Acknowledgments}
This work was supported by BMBF, HGS-hire and the Hessian LOEWE initiative through the Helmholtz International center for FAIR (HIC for FAIR). The authors thank D. Rischke, C. Greiner, M. Bleicher, J. Noronha, V. Dexheimer and P. Koch-Steinheimer for fruitful discussions. The computational resources were provided by the Frankfurt Center for Scientific Computing (CSC).

\end{document}